\documentclass[11pt]{article}

\usepackage{authblk}
\usepackage{amsmath,amssymb}
\usepackage{graphicx}
\usepackage[margin=1.25in]{geometry}
\usepackage{natbib}

\renewcommand\baselinestretch{1.5}

\newcommand{\code}{\texttt}

\title{Beyond Reproducible Research: Building a Formal Representation of a Data Analysis}

\author[1]{Roger D. Peng}
\affil[1]{Department of Statistics and Data Sciences, University of Texas at Austin}

\date{}

\begin{document}

\maketitle

\begin{abstract}
    Data analyses are often constructed in an imperative manner, where commands representing actions taken on the data are issued sequentially. The publication of these commands, along with the data, is essential to the reproducibility of the analysis by others. However, simply presenting the code and the results of running the code can hide important details about the data analyst's premises, expectations, and assumptions about the data. Understanding this analysis reasoning can be critical to evaluating the quality of an analysis and for suggesting possible improvements. We argue that a formal representation of a data analysis that externalizes its logical construction offers more useful information for statically illustrating an analyst's reasoning. Such a formal representation would allow for the evaluation of some aspects of a data analysis without the need for the data, the visualization of the logical connections leading to a conclusion, and the ability to assess the sensitivity of an analyst's assumptions to unexpected features in the data. In this paper we describe an implementation of this formal representation and how it might be applied to some common data analysis tasks.
\end{abstract}

\section{Introduction}

Statisticians often find themselves making different kinds of claims which are supported by different kinds of evidence. Scientific claims are supported by data that have been analyzed using accepted approaches and interpreted according to the standards of the discipline. For example, in epidemiology, claims about the causes of disease sometimes are accompanied by evidence following the framework of Sir Austin Bradford Hill~\citep{hill2015environment}. Statisticians also find themselves making mathematical claims describing properties of their tools and methods. These claims are supported by a different type of evidence, namely the logical proof. In both domains, the scientific and the mathematical, there are well-established approaches to presenting evidence that a particular statement is true.

Statisticians who analyze data also make a third type of claim, which we will refer to here as a data analysis statement. Data analysis statements are statements about the data at hand, as opposed to inferential statements about an unobserved population. Statements such as ``the mean of the sample was $4.6$'' or ``the simple linear regression coefficient was~$1.79$'' are examples of data analysis statements. Such statements derive directly from the dataset under analysis and, unlike inferential statements, are independently verifiable as true or false. Given that data analysis statements typically serve as the basis of broader claims, it would seem vital to have a way of evaluating the validity of the underlying data analysis statement.

In the scientific literature, data analysis statements are generally assumed to be true without the need for proof. For example, in the medical literature, it is common for the beginning of the Results section to include a recitation of summary statistics describing the study population and perhaps collect these statistics in a table~\citep{Hayes-Larson2019,yoshida:tableone:2022}. The reader is generally not given any proof or evidence that these statements about the study population are true, but rather it is assumed that whoever computed those statistics must know what they are doing and likely did the right computation. Arguably, with many data analysis statements, there is little at stake if the statement turns out to be false. But even if detailed evidence were demanded, there does not exist a universally agreed upon approach to presenting that evidence to others.


In recent decades, there has been a growing concern about the reproducibility of computational research in a variety of fields, as a number of studies have been demonstrated to state results that cannot be reconstructed from the original data~\citep{peng:2011,peng2021reproducible,leek:peng:2015,national2019reproducibility,omenn2012evolution}. In other words, the data analysis statements made in those studies were proven to be false. For example, Baggerly and colleagues spent more than 1,500 hours attempting to reproduce the results of a major computational biology study~\citep{baggerly2009deriving,coom:wang:bagg:2007}. Their investigation found numerous computational problems with the analysis, leading to the conclusion that all of the data analysis statements in the paper were false. High profile cases such as this have caused researchers in many areas to demand that published papers be accompanied by their computer code and datasets~\citep{baggerly2010disclose,haibe2020transparency}. A number of journals have implemented policies to improve the reproducibility of the papers published in their pages~\citep{peng2009reproducible,hofner2016reproducible,wrobel2024partnering}.


Inherent in the argument for publishing code and data along with a paper is that the code and the data represent a kind of ``proof'' of the data analysis statements in the paper. However, there are a few drawbacks to simply providing code and data. The first is that in the instant that an author shares the data and code, nothing has in fact been proven. An independent analyst must re-run the code on the data in order to determine whether the analysis has any issues. Depending on the complexity of the analysis, this can be a time- and resource-consuming process~\citep{willis2021trust,peng2009reproducible} and computational results must be manually linked with data analysis statements or results placed in the paper text. Even with organized computational compendia created with literate programming techniques~\citep{gentleman2007statistical}, this linkage process is not automatic. More broadly, producing the code and the data after an analysis has been done does little to prevent any sort of problem from occurring in the analysis in the first place~\citep{leek:peng:2015}. Reproducibility is only valuable for diagnosing problems on a post-mortem basis and furthermore places the burden of verification on independent third parties rather than the original authors. Finally, the data sharing requirement of reproducibility can be problematic when an analysis uses sensitive data or protected health information. Publications using such data will likely never be fully reproducible by others.

In addition to the weaknesses noted above, a limitation of the concept of reproducibility is that the analysis code alone is often insufficient for understanding the reasoning of a data analyst, therefore making it impossible to evaluate the quality of that reasoning. Statements about the data analysis are typically not contained within the code or comments and any evidence supporting those statements is not required to be included in the code for the sake of reproducibility alone. As a result, the code illustrates what was done to produce a given result, but does not provide any information about why any statement about the data might be true.

While the standard of reproducibility is a step forward in the presentation of a data analysis, it is ultimately a dynamic representation that characterizes a data analysis as a computer program to be executed rather than as a series of statements or arguments to be verified. This characterization as a computer program can lead to the erroneous conclusion that if the program runs without error and produces a sensible output, then the analysis is (in an unspecified sense) correct. But rather than reject this statement as generally not true, we could ask ourselves, ``Is it possible to design a system for writing a data analysis so that the statement is true?" Can we design a system so that the successful execution of a computer program implies a successful data analysis (by some definition)? Such a system could allow for a static formal representation (like a mathematical proof) of a data analysis that systematically presents the chain of evidence supporting a given data analysis statement. This representation could also address some of the shortcomings of the reproducible representation described above and allow for new ways to communicate analytical arguments and evaluate the quality of data analyses.

\subsection{Data Analysis as Computer Program}
\label{sec:data-analysis-program}

Consider the following pseudo-code that might be used to compute the mean of a data frame column in R.
\begin{verbatim}
dat <- read.csv("data.csv")
x <- dat[, 1]
m <- mean(x)
\end{verbatim}
From a static reading of this code, no data analysis statement can be made because we as readers do not see the output.

Suppose that an analyst ran this code and it turned out that the value of \code{m} was~$4.6$. The analyst might then make the data analysis statement, ``The mean of the first column in the data frame is $4.6$''. In this case the statement is separated from the code itself. However, the analyst could write the following code instead.
\begin{verbatim}
dat <- read.csv("data.csv")
x <- dat[, 1]
m <- mean(x)
stopifnot( 
    round(m, 1) == 4.6 
)
\end{verbatim}
We can read from the fourth expression of code that we are expecting the mean of \code{x} (rounded to one decimal place) to be $4.6$. Therefore, we can conclude that if the code ran without error, the statement about the mean of the first column would be true. There is no need to see the data in order to draw this conclusion. This version of the code informally embeds the data analysis statement in the code itself by using an assertion. 

A few things become clear with this simple example. First, this code cannot be written without seeing the data or the results of computation, i.e. it cannot be pre-specified and so it is not an analysis plan. Second, code written in this way is about describing and documenting an analysis and its expected behavior. Third, this code is not re-usable on a different dataset, i.e. it is not a computer program in the usual sense. This code is a static representation of a data analysis conducted on a specific dataset, much the same way that a data analysis statement is a statement about a specific dataset. Finally, this code still allows an independent investigator to reproduce the analysis if needed, so that the property of reproducibility is not lost when coding in this manner.

In the sections below, we will formalize the system of building a formal static representation of a data analysis and illustrate how it might be implemented in software. We show how this representation allows for externalizing the logical construction of an analysis and present a tool for visualizing the logical premises of a data analysis statement. We also show how logically constructed analyses allow for static evaluation of data analysis code and for drawing conclusions without needing to evaluate code on data.

\section{Principles for a Static Formal Representation}
\label{sec:implementing}

We present here some general principles on which we will base our implementation of a static formal representation of a data analysis. An important goal of any implementation will be to externalize the logical reasoning of the data analysis while also computing the results of the analysis itself. 

\subsection{Characterizing Data Analyses as Statements With Evidence}


If we revisit the example in Section~\ref{sec:data-analysis-program}, the original statement made about the data was ``The mean of the first column of the data frame is $4.6$''. For that simple data analysis, the following code would have been sufficient to declare the analysis reproducible.
\begin{verbatim}
compute_mean_1st_col <- function(datafile) {
    dat <- read.csv(datafile)
    mean(dat[, 1])
}
compute_mean_1st_col("data.csv")
\end{verbatim}
The structure of this code is such that given an input (the \code{data.csv} file), we produce an output (the mean of the first column). Therefore, we could make the statement, ``The \code{compute\_mean\_1st\_col()} function takes a CSV file name as input and returns the mean of the first column of the data frame." This statement can be validated easily by inspecting the R source code above. However, this statement is a programming statement (not a data analysis statement) for which the source code is the natural supporting evidence. The data analysis statement for which this code was written (``The mean of the first column of the data frame is $4.6$'') has yet to be validated by this code. In particular, if this code were to be executed without error, we would not necessarily know if the data analysis statement is true without checking that the results matched the statement.

A key challenge when applying programming paradigms to data analysis problems is the tendency to think of data analyses as taking inputs and generating outputs. This leads to a situation where the process of doing the analysis is not relevant as long as it properly produces the specified output. However, if we consider data analyses as generating data analysis statements in need of evidence, then the logical construction approach begins to make sense because it forces the checking of premises and assumptions about the data in order to produce evidence in support of the concluding statement.

\subsection{Data Analysis Statements As Class Definitions}

We can think of data analysis statements as representing the description of an object class. So the statement ``The first column of the data frame has no missing values" defines a class of objects and the creation of an object in that class is a representation of the statement being true. This method of formalization allows for writing statements about the data with a familiar programming technique. 

Having a centralized code definition associated with a given data analysis statement allows for transparency and reduces ambiguity in the meaning of the statement. Defining formal classes also allows for automatic validity checking and guaranteeing the dispatch of specific method functions. 
From a software engineering perspective, there is value in having a single definition that is applied in multiple places rather than having to run ad hoc code every time a concept is needed. From a statistical perspective, using formal definitions in a data analysis allows us to prove useful properties of the analysis using traditional logical tools~(Section~\ref{sec:static-analysis}).
Characterizing data analysis statements as class definitions ultimately means that modularity in data analysis comes with respect to the statements we can make about the data rather than in the operations (i.e. functions) we can apply to the data.


\subsection{Supporing Premises as Class Extensions}

Once a formal class have been defined to represent a specific data analysis statement, the class can be extended to include premise classes that support those statements. The use of additional premise class definitions allows for explicit specification of what kinds of evidence might support a given statement. The embedding of premises as slots in the primary class definition represents the hierarchy of evidence that ultimately points to the original data analysis statement.

The general approach for linking supporting statements to concluding statements can be written as follows. Let \code{A} and \code{B} be two different classes of objects. The statement ``\code{A} supports \code{B}'' means that 
the class definition of \code{B} has a slot for objects of class \code{A}. The advantage of this approach is that we can know from reading the class definitions that object \code{B} would only ever exist if \code{A} were a valid object and the analyst had decided that the statement encoded by class \code{A} supports the statement encoded by class \code{B}.

\section{Implementation}
\label{sec:logical}

We consider the construction of a data analysis statement resulting from a data analysis as proceeding in two phases. First, one must specify what is the data analysis statement that is intended to be made about the data. This can range in complexity from simple summaries of the data to performance metrics for complex models. In the second phase, one links the data analysis statement with supporting premises that provide evidence in support of the statement. Once we have those premises, we could think of each premise as its own data analysis statement in need of its own supporting premises. Hence, the premises may have their own premises, etc.~and it may be possible to continue backwards until we reach something external to the analysis, like an external data file, that is either trusted or otherwise verified.

\subsection{Data Analysis Statements as Classes}

In the first phase, where we specify the data analysis statement to be made, we can implement this in code by defining a class of objects that exemplify the statement. In a programming language like R, we can use the S4 class system to define a class of objects that are valid if and only if the statement we want to make is true~\citep{cham:1998}. 


\subsubsection{Example: Missing Data}
\label{sec:example-missing-data}

Consider the following statement, ``The first column of the data frame has no missing values.'' While it is unlikely that such a statement is the final result of a data analysis, it is likely that this statement might be made in the context of a broader analysis. One challenge with this statement is that there is no universal agreement over what is meant by ``missing value''. Different conventions in different disciplines can lead to a wide variety of coding of missing values. Therefore, there would be significant value in even specifying a formal definition of what is meant by ``missing value'' here. 

We can write a formal definition using an S4 class in R. Below, we specify a class of objects called \code{vector\_no\_missing}, which is meant to characterize a numeric vector that has no missing values.

\begin{verbatim}
## Definition of a numeric vector with no missing values
setClass(
    Class = "vector_no_missing",
    contains = "numeric",
    validity = function(object) {
        if(anyNA(object))
            return("NA values present")
        if(any(object == -99))
            return("-99 missing values present")
        TRUE
    })
\end{verbatim}
The \code{vector\_no\_missing} class inherits from the built-in numeric class and according to its validity method, a numeric vector is defined as having no missing values if there are no \code{NA} values and if there are no \code{-99} values. According to the mechanics of the S4 class validity method, if either \code{NA} or \code{-99} values are found in the vector, an error will be returned and the object will not be created. The following R code can then be run to read in the data from an external comma-separated-value file.
\begin{verbatim}
## Coercion method for character --> vector_no_missing
setAs("character", "vector_no_missing", function(from) {
    num <- as.numeric(from)
    new("vector_no_missing", num)
})

## Read in data and coerce columns to 'vector_no_missing' class
dat <- read.csv("data.csv", colClasses = "vector_no_missing")
\end{verbatim}
We can see from the code itself that if no error is returned, the resulting object \code{dat} is a data frame where the first column is a valid object of class \code{vector\_no\_missing}, which means there are no \code{NA} or \code{-99} values in the column (in fact, this is true for all columns in the data frame). Therefore, we do not need to run the code and see the results to determine that there are no missing values. We only need to know that the code ran without error.


\subsection{Identifying Supporting Premises}
\label{sec:supporting-premises}

Given an encoded a data analysis statement, we want to identify one or more premises (i.e. other data analysis statements) that either directly imply or at least support the conclusion. The identification of these premises can proceed directly or indirectly. In the direct approach, we simply identify statements about the data that we know support the concluding statement. For example, if we want to say that the mean of a vector is $\geq 0$, a supporting statement might be that all of the values in that vector are $\geq 0$. With the indirect approach, we consider what are the alternatives to the concluding statement being true? In other words, can we identify the various conditions under which that statement would be false? If so, then we can verify that those conditions are \textit{not} met, thereby providing evidence that the original concluding statement is true. While this approach may appear convoluted at first glance, it essentially parallels the approach we might take in a general scientific investigation: Identify possible alternative hypotheses, use the data to rule them out, thereby providing evidence in favor of your primary hypothesis. Neither the direct or indirect approach is preferred, but in some situations, one approach may be easier to implement than the other.



\subsubsection{Example: Computing the Mean of a Column}
\label{sec:example-mean}

Consider the problem of computing the mean of a column in a data frame and providing evidence that it takes a certain value. Suppose after some initial exploratory analysis, the statement we want to make is that ``The mean of the first column of the data frame is $4.6$''. Then we can first define a class of objects that are valid if that statement is true. Below we define an S4 class for this purpose.
\begin{verbatim}
setOldClass("data.frame")
setClass(
    Class = "mean_1st_col_df_is_4.6",
    slots = c(data = "data.frame"),
    validity = function(object) {
        first_col <- object@data[, 1]
        if(!isTRUE(all.equal(mean(first_col), 4.6)))
            return("mean of first column is not equal to 4.6")
        TRUE
    })
\end{verbatim}
This class in particular defines what we mean when we say ``is equal to," i.e. in the sense of \code{all.equal()}. We can now run the following code to create our newly defined object.
\begin{verbatim}
dat <- read.csv("data.csv", colClasses = "numeric")
x <- new("mean_1st_col_df_is_4.6", data = dat)
\end{verbatim}
We know by definition that \code{x} is an object that contains a data frame whose first column has a mean of $4.6$.

In many data analysis situations, we want to say something more than the basic data analysis statement at hand to indicate that there isn't anything unexpected going on with the data. The interpretation of the mean of a sample being $4.6$ might change if we knew, for example, that the distribution of the data was highly skewed to the right as opposed to symmetric. The indirect approach to identifying supporting premises can be useful in these situations.

We can ask how might the statement ``The mean of the first column of the data frame is $4.6$'' be false? We know that the \code{mean()} function can return a value that is (1)~\code{NA}, (2)~\code{Inf}/\code{-Inf}, (3)~less than $4.6$, or (4)~greater than $4.6$. If we focus on the case of \code{mean()} returning an \code{NA} value, we know that that if we can show that the first column does not have any \code{NA} values, then that would eliminate the case of \code{mean()} returning an \code{NA} value. In Section~\ref{sec:example-missing-data} we already defined a class describing a vector with no missing values. Showing that the first column of the data frame is an object of class \code{vector\_no\_missing} would imply that the mean of the first column cannot be \code{NA}. The following class characterizes the statement ``the first column of the data frame is an object of class \code{vector\_no\_missing}.''
\begin{verbatim}
setClass(
    Class = "df_1st_col_no_missing",
    slots = c(data = "data.frame"),
    validity = function(object) {
        first_col <- object@data[, 1]
        if(!is(first_col, "vector_no_missing"))
            return("first column may have missing values")
        validObject(first_col, test = TRUE)
    })
\end{verbatim}
Then we want to show that the existence of this object supports the fact that the mean of the first column is $4.6$ (in the sense that it rules out the possibility of an \code{NA} mean value). For this, we can define a new class as follows.
\begin{verbatim}
setClass(
    Class = "mean_1st_col_df_is_4.6_WITH_premise",
    contains = "mean_1st_col_df_is_4.6",
    slots = c(premise_df_1st_col_no_missing = "df_1st_col_no_missing"))
\end{verbatim}
This class extends the existing \code{mean\_1st\_col\_df\_is\_4.6} class by adding a slot that encodes the supporting premise of the first column having no missing values. Hence, the class definition encodes the statement ``If the first column of the data frame has no missing values, then the mean cannot be a missing value, and that supports the statement that the mean of the first column is $4.6$.'' We can take a similar approach to show that if the first column has no \code{Inf}/\code{-Inf} values, then the mean of that column cannot be \code{Inf}/\code{-Inf} (but we will omit this case here).

The remaining two cases are a bit more subtle than the first two because they may be interpreted in a few different ways. Consider the case of mean being greater than $4.6$. Two ways that the mean could be greater than $4.6$ are if (a)~the entire distribution were shifted over to the right, or (b)~there are some large outliers to the right-hand side of the distribution. (An analogous approach could be taken to handle the case where the mean is less than $4.6$.) There are a variety of statistical tools that could be used to investigate and dispense with these cases; here we will use a simple five-number summary. The logic we will use is that if the median of the data is close to $4.6$ (i.e. the distribution is not shifted to the left or right) and there are no significant outliers indicated by the five-number summary, then that supports the mean being $4.6$. 

First we can define the two classes that characterize the two premises that the median be close to $4.6$ and that there are no outliers.
\begin{verbatim}
setClass(
    Class = "median_close_to_4.6",
    slots = c(data = "data.frame"),
    validity = function(object) {
        first_col <- object@data[, 1]
        m <- median(first_col)
        if(abs(m - 4.6) > 0.5)
            return("median not close to 4.6")
        TRUE
    })

setClass(
    Class = "fivenum_no_outliers",
    slots = c(data = "data.frame"),
    validity = function(object) {
        first_col <- object@data[, 1]
        fn <- fivenum(first_col)
        iqr <- fn[4] - fn[2]
        if(fn[5] > fn[3] + 3 * iqr)
            return("there are outliers to the right")
        if(fn[1] < fn[3] - 3 * iqr)
            return("there are outliers to the left")
        TRUE
    })
\end{verbatim}
Note that we have explicitly defined here what we mean by ``close to 4.6'' (within $\pm 0.5$ of $4.6$) and ``significant outlier'' (more or less than $3\times IQR$ from the median). 

Now we have to re-write the original class definition for  \code{mean\_1st\_col\_df\_is\_4.6\_WITH\_premise} to accept three arguments because we now have three premises instead of one. We can do this by simply adding more slots to the original class definition.
\begin{verbatim}
setClass(
    Class = "mean_1st_col_df_is_4.6_WITH_premise",
    contains = "mean_1st_col_df_is_4.6",
    slots = c(
        premise_df_1st_col_no_missing = "df_1st_col_no_missing",
        premise_median_close_to_4.6 = "median_close_to_4.6",
        premise_fivenum_no_outliers = "fivenum_no_outliers"
    )
)
\end{verbatim}
Finally, the analysis can be executed by running the following code to show that the premises support the concluding statement.
\begin{verbatim}
dat <- read.csv("data.csv", colClasses = "vector_no_missing")
p1 <- new("df_1st_col_no_missing", data = dat)
p2 <- new("median_close_to_4.6", data = dat)
p3 <- new("fivenum_no_outliers", data = dat)
output <- new("mean_1st_col_df_is_4.6_WITH_premise",
              data = dat,
              premise_df_1st_col_no_missing = p1,
              premise_median_close_to_4.6 = p2,
              premise_fivenum_no_outliers = p3)
\end{verbatim}

We have now dispensed with the four scenarios identified in the beginning that were alternatives to the original concluding statement being true. We showed that the data do not have \code{NA} values, the median is close to $4.6$, and there are no significant outliers to the right (or left). These are all tasks that would likely be done in the normal course of a data analysis where the mean is being computed. However, we have embedded these activities in a logical framework where premises are stated in support of a concluding statement. 


\section{Static Analysis of Code}
\label{sec:static-analysis}

One benefit of our proposed representation of data analyses is that it allows for the static analysis of data analysis code, either by hand or potentially using other programs. The reason this is possible is because our approach requires that expectations, assumptions, and definitions be made explicit in the code. As a result, we can evaluate aspects of the analysis without running the code on the data, a potentially expensive or even impossible operation if data are not available. In particular, the need to create an explicitly defined class forces the analyst to externalize what it means for the analysis to be correct. Furthermore, we can use the code to draw some basic conclusions about the results of the analysis without having to see the output.

\subsection{Inferring Properties of Output}

Consider again the example in Section~\ref{sec:example-missing-data} where we had the statement ``The first column of the data frame has no missing values.'' This statement was characterized by the S4 class \code{vector\_no\_missing} (along with corresponding \code{coerce} method) so that when we called \code{read.csv()} to read in a data frame using argument \code{colClasses = "vector\_no\_missing"}, we knew that if the code ran without error, that each column of the data frame would not have any missing values (according to the given definition). If we then ran the following code under the assumption that there is at least one column in the data frame,
\begin{verbatim}
dat <- read.csv("data.csv", colClasses = "vector_no_missing")
mean(dat[, 1])
\end{verbatim}
we could be confident that the output of \code{mean()} was not \code{NA}. Hence, we can infer properties of the result without needing to see the actual results, which requires running the code.

\subsection{Avoiding Silent Errors}

Let us consider a more subtle example that is common in practice: the joining of data frames. Joining data frames can introduce pernicious errors that generate no warning if data are not formatted as expected. Consider the following two data frames:
\begin{verbatim}
  country value1 year
1      US     92 2000
2      US    117 2001
3      US     93 2002
\end{verbatim}
and
\begin{verbatim}
  country   value2 year
1     USA 48.74391 2000
2     USA 49.44440 2001
3     USA 50.21478 2002
\end{verbatim}
The goal is to join the two data frames by the \code{country} and \code{year} columns to produce a 3-row merged data frame that contains both \code{value1} and \code{value2}. It is clear that the \code{country} column in both data frames is meant to encode the same information, however in the first data frame the name \code{US} is used and in the second the name \code{USA} is used. Such minor variations in abbreviation, which are common in real datasets, can wreak havoc on joining operations. If we assume that the first data frame is stored in the file \code{data1.csv} and the second data frame in file \code{data2.csv}, then we can imagine running the following code to do a join.
\begin{verbatim}
dat1 <- read_csv("data1.csv")
dat2 <- read_csv("data2.csv")
merged_full <- full_join(dat1, dat2, by = c("country", "year"))
merged_left <- left_join(dat1, dat2, by = c("country", "year"))
merged_right <- right_join(dat1, dat2, by = c("country", "year"))
merged_inner <- inner_join(dat1, dat2, by = c("country", "year"))
\end{verbatim}
Each of the join functions produces a result that is unexpected in a slightly different way. The full join produces a data frame with 6 rows because there are no matches on the \code{country} column and \code{full\_join()} retains all rows in the original data frames. The left join produces a data frame with 3 rows but the \code{value2} column is all \code{NA}. The right join also produces a data frame with 3 rows but now with the \code{value1} column all \code{NA}. Finally, the inner join produces a data frame with zero rows. From reading the code, there is no way to know that any of these unexpected outcomes will occur without knowing what the data look like in advance. Furthermore, in all three joins, no error or warning message is produced.

The primary problem in this example is that the statement that is attempting to be made by the analyst has not been specified. As a result, there is no explicit definition of what the correct output should look like. We could encode a statement about the merged data frame using the following S4 class.
\begin{verbatim}
setClass(
    Class = "merged_dataset",
    contains = "data.frame",
    validity = function(object) {
        if(nrow(object) != 3) return("incorrect number of rows")
        if(ncol(object) != 4) return("incorrect number of columns")
        if(!all(names(object) %in% c("country", "year", "value1", "value2")))
            return("wrong column names")
        if(anyNA(object$value1) || anyNA(object$value2))
            return("NA values in 'value1' or 'value2'")
        TRUE
    }
)
\end{verbatim}
The validity method for the class defines what the correct properties of the merged data frame are, such as the number of rows and columns, the column names, and the presence of missing values. If we then run the following code,
\begin{verbatim}
dat1 <- read_csv("data1.csv")
dat2 <- read_csv("data2.csv")
merged <- new("merged_dataset",
              left_join(dat1, dat2, by = c("country", "year")))
\end{verbatim}
we know that either it will return an object of the \code{merged\_dataset} class or the validity method will return an error (in the example above an error is returned). We can predict this behavior because of the explicit class definition above.


\subsection{Lexical Analysis of Evidence and Support}

Because the relationships between statements and their supporting premises must be explicitly defined using S4 classes and methods, the code admits the possibility of these relations being examined either by hand or by some automated means. In general, if a class is defined to represent a relation between a premise and a concluding statement, we can examine the class definition to  determine the precise evidence that can support the concluding statement. This can be done by looking directly at the class definitions in the code or by using some of the tools in the \code{methods} package in R. 

For example, in Section~\ref{sec:example-mean}, we defined a class \code{mean\_1st\_col\_df\_is\_4.6\_WITH\_premise} to allow for premises for the statement ``The mean of the first column of the data frame is $4.6$''. If we wanted to know from the code what kind of evidence would support this statement, we can examine the class definition, either directly or via the \code{getClass} function.
\begin{verbatim}
> getClass("mean_1st_col_df_is_4.6_WITH_premise")
Class "mean_1st_col_df_is_4.6_WITH_premise" [in ".GlobalEnv"]

Slots:
                                                                  
Name:  premise_df_1st_col_no_missing   premise_median_close_to_4.6
Class:         df_1st_col_no_missing           median_close_to_4.6
                                                                  
Name:    premise_fivenum_no_outliers                          data
Class:           fivenum_no_outliers                    data.frame

Extends: "mean_1st_col_df_is_4.6"\end{verbatim}
From the \code{getClass} listing, we can see that one such supporting premise class is the \code{median\_close\_to\_4.6} class. Anyone interested in knowing the definition of this class could look it up directly, particularly its validity method.
\begin{verbatim}
> getClass("median_close_to_4.6")@validity
function(object) {
        first_col <- object@data[, 1]
        m <- median(first_col)
        if(abs(m - 4.6) > 0.5)
            return("median not close to 4.6")
        TRUE
    }
\end{verbatim}
Of course, running the \code{getClass} function in R requires executing some of the analysis code in R. However, we would only need to execute the class definition code in this case. We would not need to generate any objects from the classes themselves, which would require access to the data.

\subsection{Visualizing Premise Structure}

\begin{figure}[tbh]
\centering
\includegraphics[width=6in]{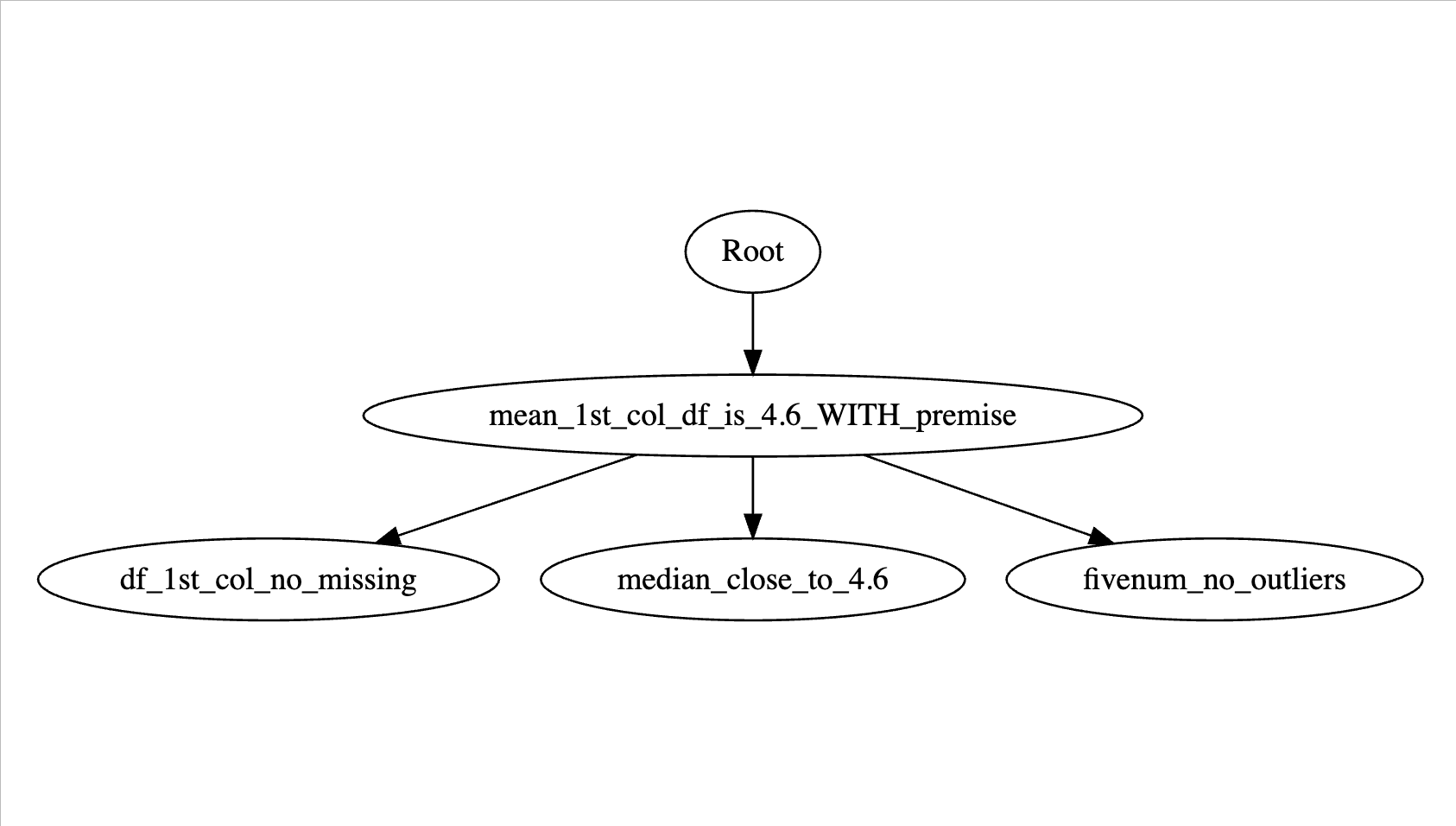}
\caption{Hierarchy of supporting premises for estimating the mean.}
\label{fig:premisetree}
\end{figure}

The output from \code{getClass} can be cumbersome if there are more than a few premises, but the hierarchical structure of the classes naturally lends itself to tree-based visualization. We have written some code to help with visualizing the premise hierarchy in a textual manner~(code listing is in Supporting Information Appendix~\ref{sec:viz-premises}). 
\begin{verbatim}
> statements("mean_1st_col_df_is_4.6_WITH_premise")
                                levelName
1 Root                                   
2  °--mean_1st_col_df_is_4.6_WITH_premise
3      ¦--df_1st_col_no_missing          
4      ¦--median_close_to_4.6            
5      °--fivenum_no_outliers   
\end{verbatim}
We have also written a corresponding \code{plot} method to visualize the tree structure of the supporting premises as a graphic~(Figure~\ref{fig:premisetree}). 

In both the textual output and the plot, the interpretation is that child nodes that are connected via a common parent node are bound together using a logical AND. So all of the premises encoded in the child nodes must be true in order for the parent node to be true. In some cases, seeing the visualization of the premise hierarchy may be sufficient for evaluating a data analysis or for identifying potential weaknesses, without any need to examine the code.

\subsection{Sensitivity Analysis}

A key advantage to implementing data analysis statements as S4 classes is that each statement is paired with a method of testing its validity via the validity method. We can then use those built-in tests to test the sensitivity of the conclusions to unexpected features in the data. These unexpected features, such as outliers or nonlinearities, could be simulated and the code could be run on these alternate datasets. A version of this approach is described in~\cite{zhang2025inside}. Here, one key goal might be to simulate data that have unexpected features but are not ``caught'' by the various validity methods in the premise statements. In such situations, one might want to add more premise statements or adjust the tolerances of the validity methods in the existing premise statements.

Consider the example from Section~\ref{sec:example-mean} where we are computing the mean of a vector. It is possible to generate data that have a mean of $4.6$ (so the concluding statement would be true) but perhaps have unexpected properties like large outliers or a skewed distribution. Simulating such data and evaluating the effectiveness of the premise statements can help ensure that if the concluding statement is made, the data have been generated in the expected manner. 

\section{Application: Simple Linear Regression}

In this section we describe a longer example of how our approach can be applied to a problem involving a simple linear regression. Often with linear regression applications, there is a coefficient of interest that is reported, perhaps representing the association between some exposure of interest and an outcome. At the same time, merely stating the coefficient (and its uncertainty) is generally not the only statement an analyst will make. Rather, an analyst will want to demonstrate that there is nothing unusual or unexpected occurring in the data and that we have some reasonable understanding of the data generation process. Making such a composite statement will typically involve a number of checks on the data, particularly of the model residuals.

For the purposes of this demonstration we have simulated some bivariate data with a simple linear regression slope of $1.79$. In an analysis of this dataset, a statement we might like to make is that ``the slope of the regression line is $1.79$''. Such a statement could be encoded in the following class.
\begin{verbatim}
setClass(
    Class = "slr_slope_1.79",
    slots = c(
        fit = "lm"
    ),
    validity = function(object) {
        beta <- round(coef(object@fit), 2)
        if(beta[2] != 1.79)
            return("slope coefficient does not match claim")
        TRUE
    }
)
\end{verbatim}

In addition to this primary statement, we would also like to provide supporting premises. These premises would involve statements regarding 
(1)~the lack of any nonlinearity in the residuals; (2)~the lack of significant outliers in the response or in the predictor; and (3)~the appropriateness of certain plots of the residuals. These premises could be added by extending the class above in the following manner.
\begin{verbatim}
setClass(
    Class = "slr_slope_1.79_WITH_premise",
    contains = "slr_slope_1.79",
    slots = c(
        premise_slr_no_nonlinearity = "slr_no_nonlinearity",
        premise_slr_no_outliers = "slr_no_outliers",
        premise_slr_plots_look_okay = "slr_plots_look_okay"
    )
)
\end{verbatim}
The three premise classes indicated in the class definition still need to be defined, and we show the complete implementation in Supporting Information Appendix~\ref{sec:slr-example}. For checking nonlinearity, we can define the following class, which simply checks for possible quadratic structure in the data. 
\begin{verbatim}
setClass(
    Class = "slr_no_nonlinearity",
    slots = c(
        fit = "lm"
    ),
    validity = function(object) {
        fit_alt <- lm(y ~ poly(x, degree = 2),
                      data = model.frame(object@fit))
        llr <- as.numeric(logLik(fit_alt) - logLik(object@fit))
        if(llr >= 7)
            return("strong evidence of nonlinearity in the data")
        TRUE
    }
)
\end{verbatim}
Here, if the log-likelihood ratio between the quadratic model and the original linear model is greater than $7$ (i.e. very strong evidence of quadratic structure), the validity method returns an error.

\begin{figure}[tbh]
\centering
\includegraphics[width=6in]{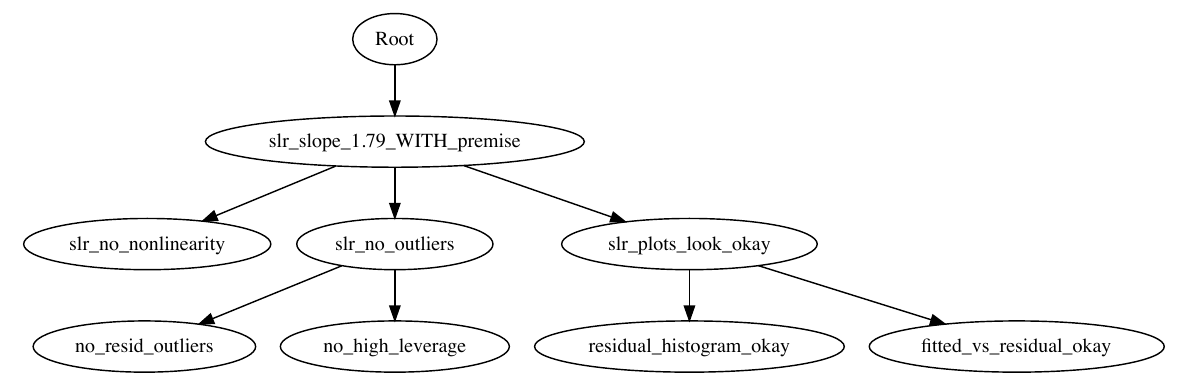}
\caption{Hierarchical structure of premises for simple linear regression statement.}
\label{fig:slr-tree}
\end{figure}

One of the premises involves a check of the residuals using two standard residual plots: a histogram of the standardized residuals and a scatterplot of the fitted values versus the residuals.
\begin{verbatim}
setClass(
    Class = "slr_plots_look_okay",
    slots = c(
        premise_residual_histogram_okay = "residual_histogram_okay",
        premise_fitted_vs_residual_okay = "fitted_vs_residual_okay"
    )
)
\end{verbatim}
Again, the two premise classes here still need to be defined. We show one of the below. Given that it is not straightforward to automate checking of visualizations, we take a manual approach here and force the analyst to view the plot and interactively confirm that it looks as-expected (by whatever definition the analyst finds appropriate). Below, we show how we implement a check of the histogram of the standardized residuals. Once the analyst confirms that the histogram is acceptable, the validity method returns. Otherwise, an error is returned.
\begin{verbatim}
setClass(
    Class = "residual_histogram_okay",
    slots = c(
        fit = "lm"
    ),
    validity = function(object) {
        r <- local({
            res <- rstandard(fit)
            hist(res, main = "Standardized Residuals")
            readline("Does the histogram look okay? [y/n] ")
        })
        if(tolower(r) != "y")
            return("problem with histogram of standardized residuals")
        TRUE
    }
)
\end{verbatim}
The full tree of premises (Figure~\ref{fig:slr-tree}) can be visualized using the plot method implemented in Supporting Information Appendix~\ref{sec:viz-premises}.

In contrast to our approach, a traditional analysis plan for a simple linear regression problem might look as follows. 
\begin{verbatim}
## Fit model and check coefficients
fit <- lm(y ~ x, dat)
coef(fit)

## Diagnostic plots
hist(rstandard(fit))
plot(fitted(fit), resid(fit))

## Check leverages and residuals
summary(hatvalues(fit))
summary(rstandard(fit))

## Check for quadratic nonlinearity
fit_alt <- lm(y ~ poly(x, degree = 2), dat3)
logLik(fit_alt) - logLik(fit)
\end{verbatim}
While the execution of methods is largely equivalent to our approach, what is missing from this code listing is any understanding of what is expected by the analyst. Simply knowing that the above code executed without error tells us little about the analysis, except perhaps that the usual tools were applied.

\section{Discussion}
\label{sec:discussion}

We have presented a new approach to representing a data analysis that externalizes analysts' assumptions and expectations about the data in a transparent manner that others can evaluate, interpret, and criticize. Our approach represents a data analysis as a logical construction of data analysis statements that support each other in a hierarchical manner. We build on the ideas behind reproducible research~\citep{peng:domi:zege:2006,peng:2011}, whereby code is provided for merely reproducing the outputs, and add a semantic layer on top of the code so that independent analysts can understand the meaning of what was intended by the code. Our approach implements data analysis statements as class definitions where premise classes indicate the relationships between different data analysis statements. As shown in the examples, a key advantage of this approach to communicating a data analysis is that it forces the analyst to be explicit about what is the expected result. This explicit representation allows for a static analysis of the code and the quality of the analysis itself. Unlike with a reproducible analysis, a logically constructed analysis does not necessarily require executing the code on the data in order to understand important characteristics of the analysis.




Aspects of this representation of a data analysis borrows from the propositions as types principle from computer science~\citep{harper2016practical}. There, propositions are considered as types in a programming language and the existence of a term of that type is considered proof that the proposition is true. Here, we use the class system in R to characterize a data analysis statement and the existence of an object of that class as an indication that the statement is true. Furthermore, we add a system for developing and characterizing supporting premises, which is important if data analyses are to be used to draw stronger conclusions. Like with the propositions as types principle, we consider the classes representing data analysis statements as regular objects in the programming language. These objects can be manipulated like any other object and therefore the statements they encode could be re-used to support other data analyses.

The work described here also has some relationship with the concept of unit testing~\citep{zhu1997software,wickham2023r,zhang2025inside}. In general, unit testing can be used to ensure that software or functions that are written meet a certain specification by checking that specific inputs match expected outputs. The key requirement for unit testing to make sense is a clear understanding of what the expected output is for a given function. Similar to our logically constructed data analyses, the clear specification required of unit testing makes it a successful tool for software development. However, the context in which unit testing is employed is different from what we describe in this paper. Rather than ensure that a software package meets a given specification, we aim to represent a data analysis argument through a series of connected data analysis statements.


A clear question with our framework is how best to incorporate graphical summaries and methods. In general, it is not clear what is the optimal way to express one's expectation for a graphical summary without first making the graphic. Some work has been done along these lines to facilitate graphical comparisons such as unit testing using raster images and pixel-by-pixel comparisons~\citep{murrell:gdiff:2023}. Textual formats for images, such as the scalable vector graphics (SVG) format, allow for the application of textual difference tools to assess variation~\citep{henry:vdiffr:2024}. However, these approaches, and related ones like image hashing~\citep{swaminathan2006robust}, all require that the image be created first as a basis for comparison. The problem of formalizing in code the typically intuitive process of assessing whether a visual summary meets expectations or not is likely a challenging one and in need of further work.


An interesting connection arises between the hierarchical trees that we use to visualize our data analysis premises and the world of systems engineering. The trees shown in Figures~\ref{fig:premisetree} and~\ref{fig:slr-tree} are related to the concept of a fault tree, which is a tool commonly used in systems engineering for conducting a structured risk assessment and has a long history in aviation, aerospace, and nuclear power applications~\citep{vesely1981fault,michael2002fault}. Fault trees are esseentially the logical negation of the trees found in Figures~\ref{fig:premisetree} and~\ref{fig:slr-tree}. Whereas our trees indicate what must be true in order for the top level statement to be true (i.e.~a success), fault trees indicate what must be tree for an \textit{unexpected} outcome to occur (i.e.~a failure). In a sense, they are related to the indirect approach we describe for identifying supporting premises in Section~\ref{sec:supporting-premises}. Fault trees are well-established in the systems engineering literature and yet we struggle to find any explicit use of this concept in the statistical analysis literature. In the development of the logical structure of a data analysis, they may be useful for identifying supporting premises for a data analysis statement by providing scenarios that can be checked with the data.


The S4 class and method system in R is used here to provide concrete examples of how this logical construction could be implemented. It is clear that this specific implementation is too verbose to be practical in real-world data analyses. The creation of classes for every statement generates substantial coding overhead. However, it is worth noting that much of that additional coding is dedicated to specifying the definitions of what we expect to observe in the output. While a simple analysis might be reproducible with a few lines of code, we see from the examples here that there is substantial information that is omitted in such a presentation. Making that information explicit, whether using S4 classes or some other approach, will require more code. That said, it is likely that a more efficient system could be designed to communicate this information and adding such a system to R may be a fruitful area of future work. 

Ultimately, what we propose is a formal representation for what was done in a data analysis, not a tool for doing a data analysis. As such, we do not suggest that such a representation inherently improves the quality of an analysis. However, if we are to improve the general quality of data analyses, a useful tool would be to have a clear representation of the analyst's thinking and reasoning. We believe that our framework allows for that reasoning to be made more transparent than with current approaches that focus on reproducibility.






\bibliographystyle{asa}
\bibliography{combined}

\clearpage

\appendix

\renewcommand{\baselinestretch}{1}

\section*{Supporting Information}

\section{Visualizing Supporting Premises}
\label{sec:viz-premises}

\begin{verbatim}
setClass(
    Class = "statement_object",
    slots = c(
        output = "list"
    )
)

## Looks up premises based on class name
get_premises <- function(clname) {
    def <- getClass(clname)
    s <- slotNames(def)
    m <- grepl("^premise", s)
    if(!any(m)) {
        return(list(NULL))
    }
    else {
        premises <- lapply(s[m], function(nm) def@slots[[nm]])
        names(premises) <- unlist(premises)
        lapply(premises, get_premises)
    }
}

## Organize premises into a hierarchically organized list
statements <- function(x) {
    if(isS4(x))
        cl <- class(x)
    else if(is.character(x))
        cl <- x
    else
        stop("cannot handle object")
    pr <- get_premises(cl)
    output <- list(pr)
    names(output) <- cl
    new("statement_object", output = output)
}

## Print tree structure of premise statements
setMethod(
    "show",
    "statement_object",
    function(object) {
        tree <- data.tree::as.Node(object@output)
        print(tree)
        invisible(tree)
    }
)

## Plot tree structure of premise statements
setMethod(
    "plot",
    "statement_object",
    function(x, y, ...) {
        tree <- data.tree::as.Node(x@output)
        plot(tree, ...)
    }
)
\end{verbatim}

\section{Classes For Simple Linear Regression Example}
\label{sec:slr-example}

\begin{verbatim}
setClass(
    Class = "slr_slope_1.79",
    slots = c(
        fit = "lm"
    ),
    validity = function(object) {
        beta <- round(coef(object@fit), 2)
        if(beta[2] != 1.79)
            return("slope coefficient does not match claim")
        TRUE
    }
)

setClass(
    Class = "slr_no_nonlinearity",
    slots = c(
        fit = "lm"
    ),
    validity = function(object) {
        fit_alt <- lm(y ~ poly(x, degree = 2),
                      data = model.frame(object@fit))
        llr <- as.numeric(logLik(fit_alt) - logLik(object@fit))
        if(llr >= 7)
            return("strong evidence of nonlinearity in the data")
        TRUE
    }
)

setClass(
    Class = "slr_no_outliers",
    slots = c(
        premise_no_resid_outliers = "no_resid_outliers",
        premise_no_high_leverage = "no_high_leverage"
    )
)

setClass(
    Class = "no_resid_outliers",
    slots = c(
        fit = "lm"
    ),
    validity = function(object) {
        res <- rstandard(object@fit)
        if(any(abs(res) > 4))
            return("some standardized residuals are greater than +/-4")
        TRUE
    }
)

setClass(
    Class = "no_high_leverage",
    slots = c(
        fit = "lm"
    ),
    validity = function(object) {
        h <- hatvalues(object@fit)
        if(any(h > 5 * mean(h)))
            return("some points have very high leverage")
        TRUE
    }
)

setClass(
    Class = "slr_plots_look_okay",
    slots = c(
        premise_residual_histogram_okay = "residual_histogram_okay",
        premise_fitted_vs_residual_okay = "fitted_vs_residual_okay"
    )
)

setClass(
    Class = "residual_histogram_okay",
    slots = c(
        fit = "lm"
    ),
    validity = function(object) {
        r <- local({
            res <- rstandard(fit)
            hist(res, main = "Standardized Residuals")
            readline("Does the histogram look okay? [y/n] ")
        })
        if(tolower(r) != "y")
            return("problem with histogram of standardized residuals")
        TRUE
    }
)

setClass(
    Class = "fitted_vs_residual_okay",
    slots = c(
        fit = "lm"
    ),
    validity = function(object) {
        r <- local({
            g <- fit |>
                augment() |>
                ggplot(aes(.fitted, .resid)) +
                geom_point() +
                geom_hline(yintercept = 0, lty = 2) +
                labs(x = "Fitted values", y = "Residuals")
            print(g)
            readline("Does this residual plot look okay? [y/n] ")
        })
        if(tolower(r) != "y")
            return("problem with residuals vs. fitted plot")
        TRUE
    }
)

setClass(
    Class = "slr_slope_1.79_WITH_premise",
    contains = "slr_slope_1.79",
    slots = c(
        premise_slr_no_nonlinearity = "slr_no_nonlinearity",
        premise_slr_no_outliers = "slr_no_outliers",
        premise_slr_plots_look_okay = "slr_plots_look_okay"
    )
)
\end{verbatim}

\end{document}